%% file: driver.tex
\documentclass[final,1p,times]{elsarticle}

\usepackage{pifont}
\usepackage{natbib}
\usepackage{geometry}
\usepackage{graphicx}
\usepackage{txfonts}
\usepackage{subfigure}
\usepackage{wrapfig}
\usepackage{color}
\usepackage{booktabs}
\usepackage{caption}
\usepackage{amssymb}
\usepackage{url}
\usepackage{colortbl}
\usepackage{multirow}
\usepackage{fnbreak}
\usepackage{helvet}
\usepackage[font=bf]{caption}

\graphicspath{{figs/}}

\definecolor{Gray}{gray}{0.7}

\begin{document}

\begin{frontmatter}

\begin{center}
\small \textbf{9\textsuperscript{th} IAA Planetary Defense Conference -- PDC 2025} \\ \textbf{5--9 May 2025, Stellenbosch, Cape Town, South Africa} \\ \vspace{0.1in} \textbf{IAA-PDC25-04-204} \\
\end{center}

\title{NEO and imminent impactor discoveries from Hungary: recent results and lessons learnt}  


\author[fml1,fml4]{Norton O. Szab\'o\corref{cor1}}
\ead{szabo.norton@csfk.org}
\author[fml1]{Kriszti\'an S\'arneczky\corref{cor1}}
\ead{sarneczky.krisztian@csfk.org}
\author[fml1]{L\'aszl\'o L. Kiss\corref{cor1}}
\ead{kiss.laszlo@csfk.org}
\author[fml2]{Szabolcs Velkei}
\ead{szabolcs.velkei@mi.services}
\author[fml1,fml3]{Attila B\'odi}
\author[fml1,fml4]{Zs\'ofia Bora}
\author[fml1]{Bal\'azs Cs\'ak}
\author[fml1]{Borb\'ala Cseh}
\author[fml1,fml4]{\'Agoston Horti-D\'avid}
\author[fml1,fml4]{Andr\'as Jo\'o}
\author[fml1,fml4]{Csilla Kalup}
\author[fml1]{Zolt\'an Kuli}
\author[fml1]{L\'aszl\'o M\'esz\'aros}
\author[fml1,fml4]{Andr\'as P\'al}
\author[fml1]{B\'alint Seli}
\author[fml1]{\'Ad\'am S\'odor}
\author[fml1]{R\'obert Szak\'ats}
\author[fml1,fml4]{N\'ora Tak\'acs}

\cortext[cor1]{Corresponding authors: L\'aszl\'o L. Kiss, Kriszti\'an S\'arneczky, Norton O. Szab\'o}

\address[fml1]{Konkoly Observatory, HUN-REN Research Centre for Astronomy and Earth Sciences, H-1121 Budapest, Konkoly Th.M. 15-17, Hungary}
\address[fml2]{{Machine Intelligence Inc, H-2015 Szigetmonostor, Horánygyöngye 102/b, Hungary}}
\address[fml3]{Department of Astrophysical Sciences, Princeton University, 4 Ivy Lane, Princeton, NJ 08544, USA}
\address[fml4]{ELTE Eötvös Loránd University, Institute of Physics and Astronomy, Budapest, Hungary}

\input{abs}

\input{keywords}

\end{frontmatter}

%
%
\input{s-intro}

\input{s-sec1}

\input{s-sec2}

\input{s-sec3}

\input{s-conc}

\input{ack}

\bibliographystyle{model1-num-names}   
\bibliography{references}             

\vspace{0.25in}

%
%

\end{document}

%% file: abs.tex
\begin{abstract}
2022 EB5, 2023 CX1 and 2024 BX1: these are the three recent imminent impactor discoveries from the Piszk\'estet\H{o} Mountain Station of the Konkoly Observatory. They make up about one percent of all NEO discoveries from our observatory and here we provide a detailed description of our approach and methodology that led to this noticeable observational sensitivity to these meter-sized impactors. After outlining the historical background of astronomical discoveries from Hungary, we introduce our recently upgraded survey instrumentation and outline the observational strategy and its implementation. We highlight the importance of strong feedback between analysis and ongoing data collection, maximizing the value of immediate follow-up. Finally, we discuss plans for moving forward to increase the sensitivity and the temporal coverage of our survey.
\end{abstract}

%% file: keywords.tex
\begin{keyword}
Near-Earth Objects \sep imminent impactors \sep NEO survey \sep Piszk\'estet\H{o} 
\end{keyword}

%% file: s-intro.tex
\section{Introduction}

\noindent Astronomical discoveries from Hungary -- such as the discovery of new asteroids, comets, supernovae, and other transient objects -- have always been intimately connected with the continuous development of the Konkoly Observatory in Budapest, founded as state-funded astronomical institute in May 1899. Systematic monitoring of the transient phenomena and moving sky objects began in the 1930s, when Gy\"orgy Kulin (1905-1989), a very enthusiastic observer with the 24$^{\prime\prime}$ telescope located in the Budapest headquarter of the institute discovered 21 main-belt asteroids and one comet between 1936 and 1942 \cite{kulin1939, vargha1999}.

The second wave of discoveries came with the opening of the Piszk\'estet\H{o} Mountain Station of the Konkoly Observatory (MPC Observatory code: K88) in the M\'atra mountains, approximately 80 kms north-east of the capital (19.89367° E, 47.917455° N, 983.5 m) with 160-200 clear nights when observations can be made each year. Here the first telescope to begin scientific operations in 1962 was a 60/90/180 cm Schmidt telescope from Carl Zeiss Jena, which allowed capturing a 5-degree circular field-of-view (FoV) on 16$\times$16-cm glass photographic plates. A systematic search for extragalactic supernovae started immediately and the most active observer was Mikl\'os Lovas (1931-2019). While his primary interest lay in supernovae, and hence the inventory of his discoveries is dominated by 42 of them found between 1963 and 1995, Lovas is also well-known for his five comets and one NEO \cite{vargha1999, Sarneczky2019Elment}. The latter is 1982~BB =(3103) Eger, a high-albedo minor planet suspected to be related to the sources of the aubrite meteorites \cite{Eger_aubrite, bischoff2024aubrite, jenniskens2025aubrite}. 

\begin{figure*}[t!]
    \begin{center}
        \includegraphics[width=\textwidth]
        {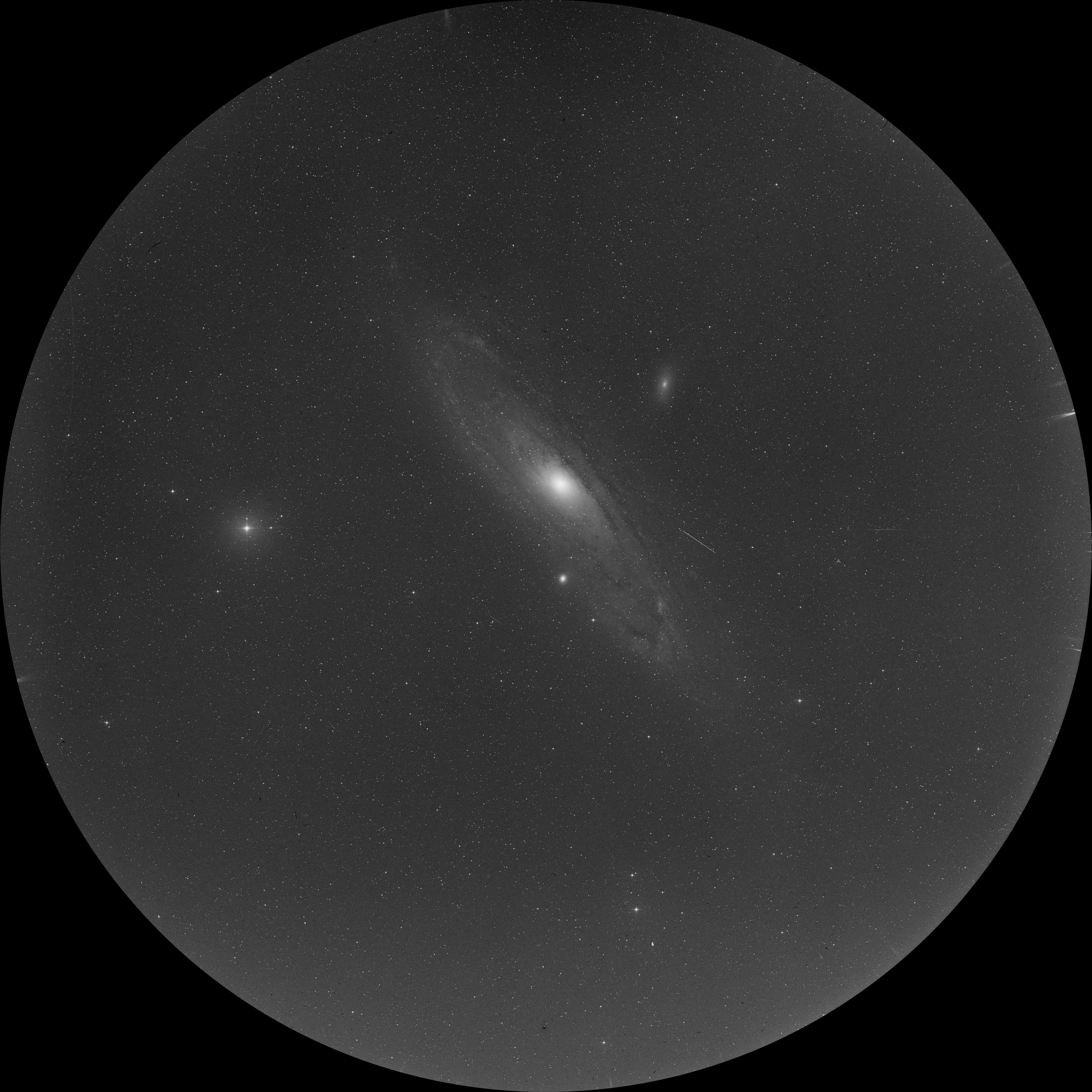}
        \caption{\small{A pre-discovery image of (3200) Phaethon, the parent body of the Geminids meteor shower. Here we reproduce the full 5-degree circular field captured by Mikl\'os Lovas on 15 December 1974, nine years before the official discovery using the IRAS satellite. Phaethon's trailed image is located slightly offset from the center to the right \textbf{\textit{next to M31}}.}}
        \label{fig:fig1}
    \end{center}
\end{figure*}

It is a recently uncovered and hence less known fact that Lovas had a rare missed opportunity on 15 December 1974, when he recorded a strange streak near the Andromeda nebula (Fig.~\ref{fig:fig1}). After the full digitization of approximately 13,000 Schmidt plates taken between 1962 and 1997 was carried out, co-author KS in 2018 made the connection between the mysterious Lovas observation and the Apollo-type Near-Earth asteroid (3200) Phaethon, which was discovered in 1983 \cite{Phaethon_disc}, nine years after the by Lovas. Having been recorded on only two photographic plates and noted only days after the observations, Lovas had never had a chance to recover the fast-moving object. The two scanned fits files\footnote{The full digitized Schmidt plate archive is available at {\tt https://schmidt-heritage.konkoly.hu/} \cite{Schmidt_her}} were used to approximate the coordinates of the previously unidentified moving object, which was thus identified with the parent body of the Geminids meteor shower \cite{Phaethon_gem}. 

With the advent of the CCD era, photographic observations quickly went out of fashion. At Piszk\'estet\H{o}, the first CCD in the Schmidt was installed in 1997 and in the following year KS and LLK started the initial continuous asteroid astrometry project, known as the JATE Asteroid Survey (JAS)\footnote{{\tt https://titan.physx.u-szeged.hu/$\sim$sky/jas/}}. Since then the project has developed both in scope and instrumentation, which finally led to the current status of running a full-time dedicated NEO survey with the recently upgraded Schmidt-telescope.

%% file: s-sec1.tex
\section{The CCD evolution at Piszk\'estet\H{o}}

\noindent It is interesting to recall how the efficiency of discovering NEOs from Piszk\'estet\H{o} evolved with the continuous upgrades of the CCD in the focal plane of the Schmidt-telescope. Between 1997 and 2010 the main imager was a Photometrics AT200 CCD camera (1536$\times$1024 KAF 1600 MCII coated CCD chip), with which the projected sky area was 29$^\prime\times$18$^\prime$, corresponding to an
angular resolution of 1 $^{\prime\prime}$/pixel. With this setup, the JAS program discovered several hundred main-belt asteroids and one NEO (2008 UZ201) over a period of 14 years.

The first upgrade was done in 2010, when a 4k$\times$4k CCD was installed, delivering a total FoV of 70$^\prime\times$70$^\prime$. The asteroid survey continued with approximately one week per month observing time on the Schmidt. Between 2010 and 2019, these efforts led to the discovery of several thousand main-belt asteroids and seven NEOs in total. While the discovery statistics showed some improvement with the larger FoV, it was clear that the full advantage of the Schmidt optics must be used to maximize the discovery rate. 

In 2020, the Schmidt went through the most extensive refurbishment and upgrade to date. Using a custom-built field flattener lens as the camera window, a new imager was installed. The camera is a back-illuminated STA 1600LN 10k$\times$10k CCD that provides a full imaging area of 95mm$\times$95mm. This translates to about 9 square degrees FoV, with the same 1 $^{\prime\prime}$/pixel image scale as with the previous two CCDs. We use a broadband Pan-STARRS w filter with a near-IR cutoff, so that we use the whole optical range to find the faint moving targets. Since the instrument upgrade the Piszk\'estet\H{o} NEO survey has been using 100\% of the Schmidt telescope time.

The regular observations are taken in remote access mode on all clear or partially clear nights, except around the full Moon $\pm$3 days, when measurements are only made when the
atmosphere is completely cloud-free, dry and calm. The decision on taking data is done by KS, based on the remotely available weather and atmospheric data collected by various sensors, such as all-sky cameras, weather stations, and raw images with the Schmidt.
The observations are controlled by the
{\tt ccdsh} environment, an in-house developed front-end control system based on various utilities including the {\tt fitsh} package \cite{Fitsh}. We take 12 to 18 sidereal tracking
images per pointing, with exposure times varying from 9 to 26 seconds, depending on the background brightness, transparency and seeing of the sky.

The images are then processed by Tycho Tracker \cite{Tycho_tracker}, using the synthetic tracking method. Four GPU cards are used for the computations, typically 0-360 deg PA range, and faster than 2"/min angular speed, with the granularity of 67\%. The maximum angular speed depends on the length of exposure, which is usually 20"/min, but can be as high as 45"/min for short exposures.

The completed 12-18 images series is processed and NEO candidates are identified by the observer until the next series is completed. This way, we already know whether there are any new NEOs in the images up to 12 minutes after the last exposure in the series. To extend the discovery space, for the past year we have been testing a new, AI-based software tool called NEODetect to identify fast-moving minor planets, which are missed completely by the traditional methods as the appear as streaks instead of moving dots on observations (for details see our companion paper by \cite{piszkes_neo_ai}, this volume). Regardless of the discovery method, the coordinates of the new NEOs are immediately reported to the MPC, and the search is interrupted to start taking follow-up images.

%% file: s-sec2.tex
\section{Survey Performance and Statistics} 

\begin{figure}[h!]
    \centering
    \includegraphics[width=\textwidth]{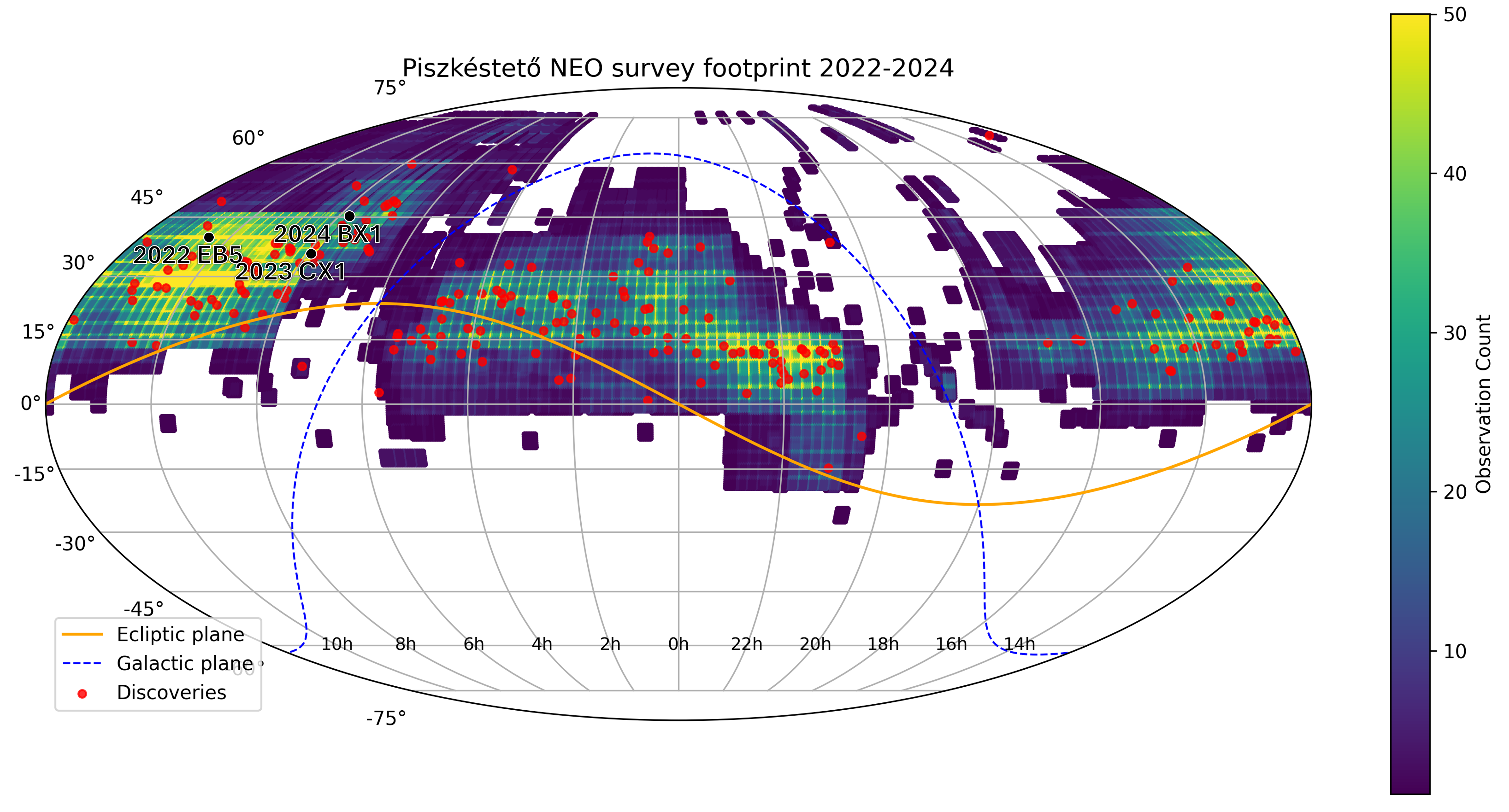}
    \caption{The Piszk\'estet\H{o} NEO survey and the discoveries between 2022-2024. \textbf{\textit{The three imminent impactor discovered by our survey are marked and labeled.}}}
    \label{fig:footprint}
\end{figure}

\begin{figure}[h!]
    \centering
    \includegraphics[width=\textwidth]{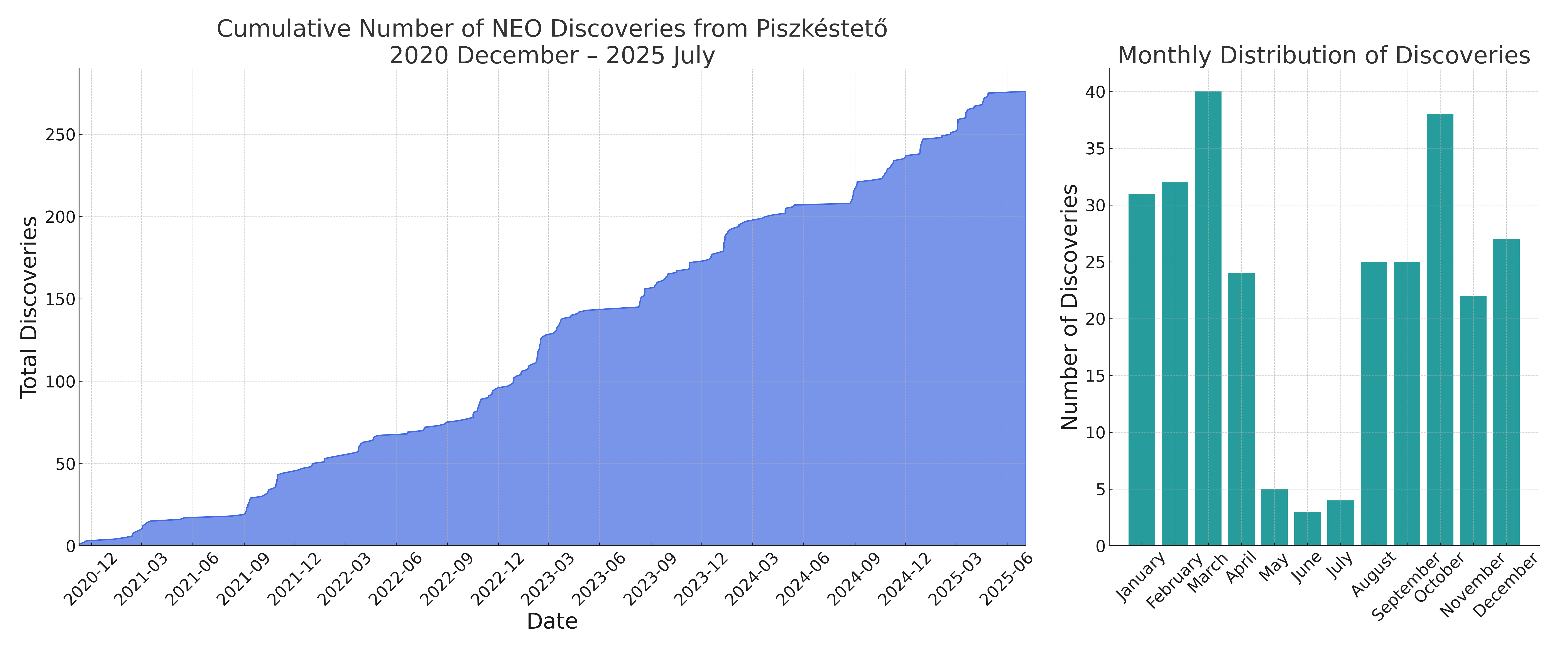}
    \caption{Cumulative and monthly distribution of NEO discoveries from Piszkéstető between December 2020 and July 2025. The left panel shows the cumulative growth in total discoveries, while the right panel illustrates the average monthly distribution across the same period. Discovery activity is consistently lower during summer months (May–July), due to short nights and reduced observing time, while a marked increase follows each autumn as night length and observing opportunities improve.}
    \label{fig:cumulative}
\end{figure}

\noindent As of July 2025, the total number of our NEO discoveries reached 289. 97\% of our discoveries are confirmed as we make immediate follow-up observations on our NEOs arter discovery. Our numbers keep increasing every week by one or two new NEOs from the autumn to spring periods as seen in Fig.~\ref{fig:cumulative}. This is a tremendous change  compared to the past. The first-light observations of the STA-1600 LN CCD were obtained in August 2020. The year after was spent on testing several approaches and observing strategies until we ended up with the one described in the previous section. To see the annual statistics and possible trends, we took the three full years of standard operations, that is from 2022 to 2024.

First we show the full footprint of the Piszk\'estet\H{o} NEO survey for the whole three years in Fig.~\ref{fig:footprint}. The 3$\times$3 deg fields are color-coded by the total number of visits. The mid-northern latitude of Piszk\'estet\H{o} (+47.9 deg) explains the dominance of the northern sky in the footprint; the NEO search rarely goes south of the ecliptic plane and the most southern parts of the ecliptic are completely left out. The galactic plane is also avoided for most of the time, although one NEO was once discovered in the dark dust lanes in the middle of the Cygnus Milky Way, where the dust obscuration of stars makes the discovery of moving objects possible despite the low galactic latitude. 

To quantify the observations: the total sky coverage between 2022 and 2024 was 17,290 square degrees (about 42\% of the whole sky), with typical observation counts between 20 to 50 (one field is counted only once on a given night even if it was revisited on the same night). Observations are archived, and are revisited for precoveries in case other surveys find objects missed by us. The annual changes reflect the weather pattern: 

\begin{itemize}
    \item 2022: 11,577 square degrees;
    \item 2023: 9,480 square degrees;
    \item 2024: 8,791 square degrees.
\end{itemize}

From these results, approximately 10,000 square degrees of sky coverage can be expected in a typical year, corresponding to about 25\% of the entire sky.

\begin{figure}[h!]
    \centering
    \includegraphics[width=1\linewidth]{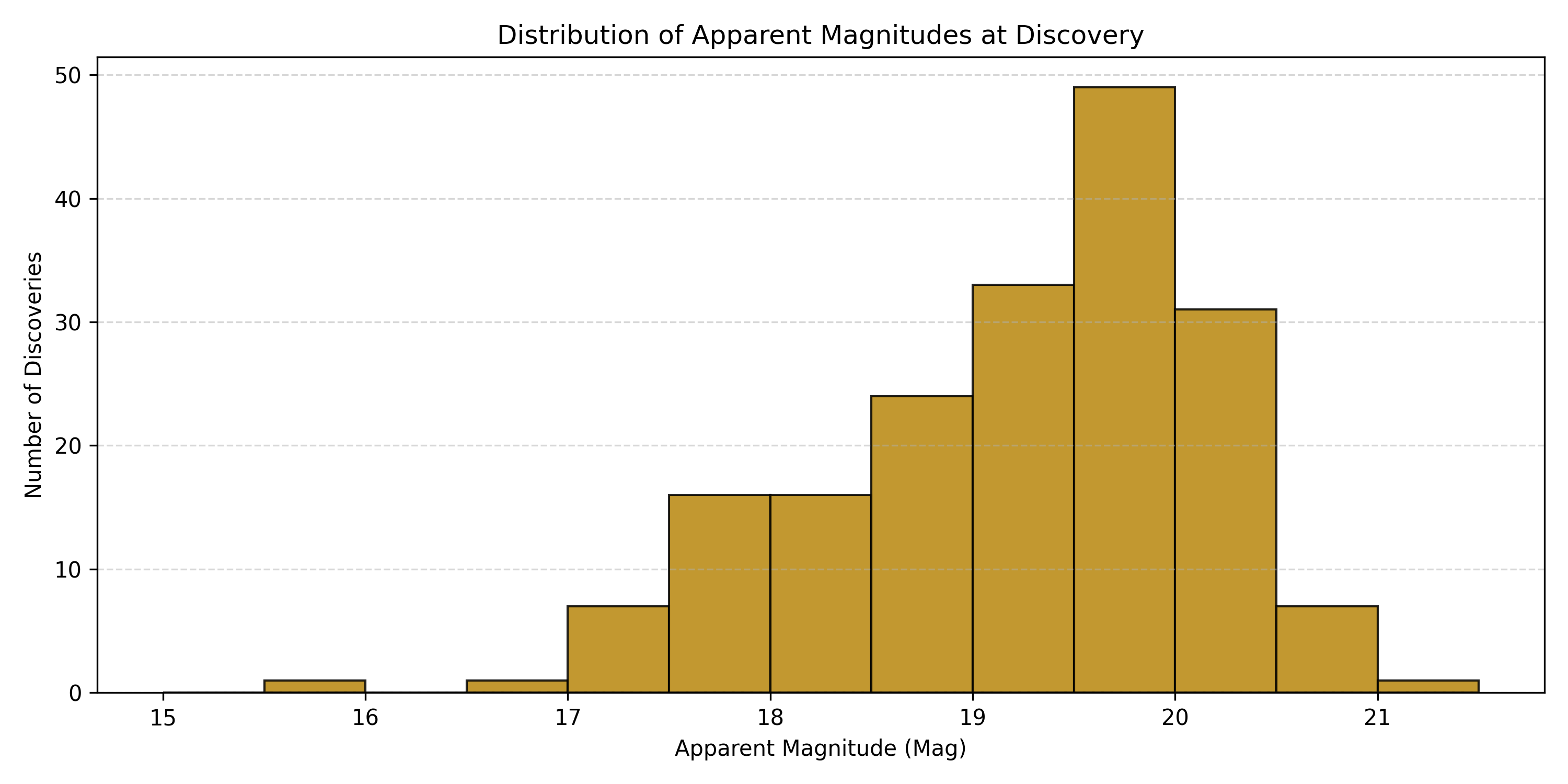}
    \caption{The apparent magnitude distrubution (Gaia G magnitudes) of the Piszk\'estet\H{o} NEOs between 2022 and 2024.}
    \label{fig:apparent-mag}
\end{figure}

The sensitivity of the Piszk\'estet\H{o} NEO survey can be characterized by two distributions. The first is the one that shows the apparent magnitude distribution of the NEO discoveries, plotted in Fig.~\ref{fig:apparent-mag}. While there were a few discoveries in the bright range from 15 to 17 mag (usually during bright nights close to the full Moon), most of the new objects fall between magnitude 18-20, with a steep decline towards the faint limit that is quite sharp at magnitude 21. (Note that our magnitude values are derived from field stars with Gaia G magnitudes, while the actual observations are done through a broad-band Pan-STARRS w filter with IR cutoff.) From the shape of the distribution one can conclude that our discoveries are close to being complete down to mag. 19-19.5, the maximum of the skewed distribution, while for the fainter targets we loose the sensitivity quite drastically. 

\begin{figure}[h!]
    \centering
    \includegraphics[width=1\linewidth]{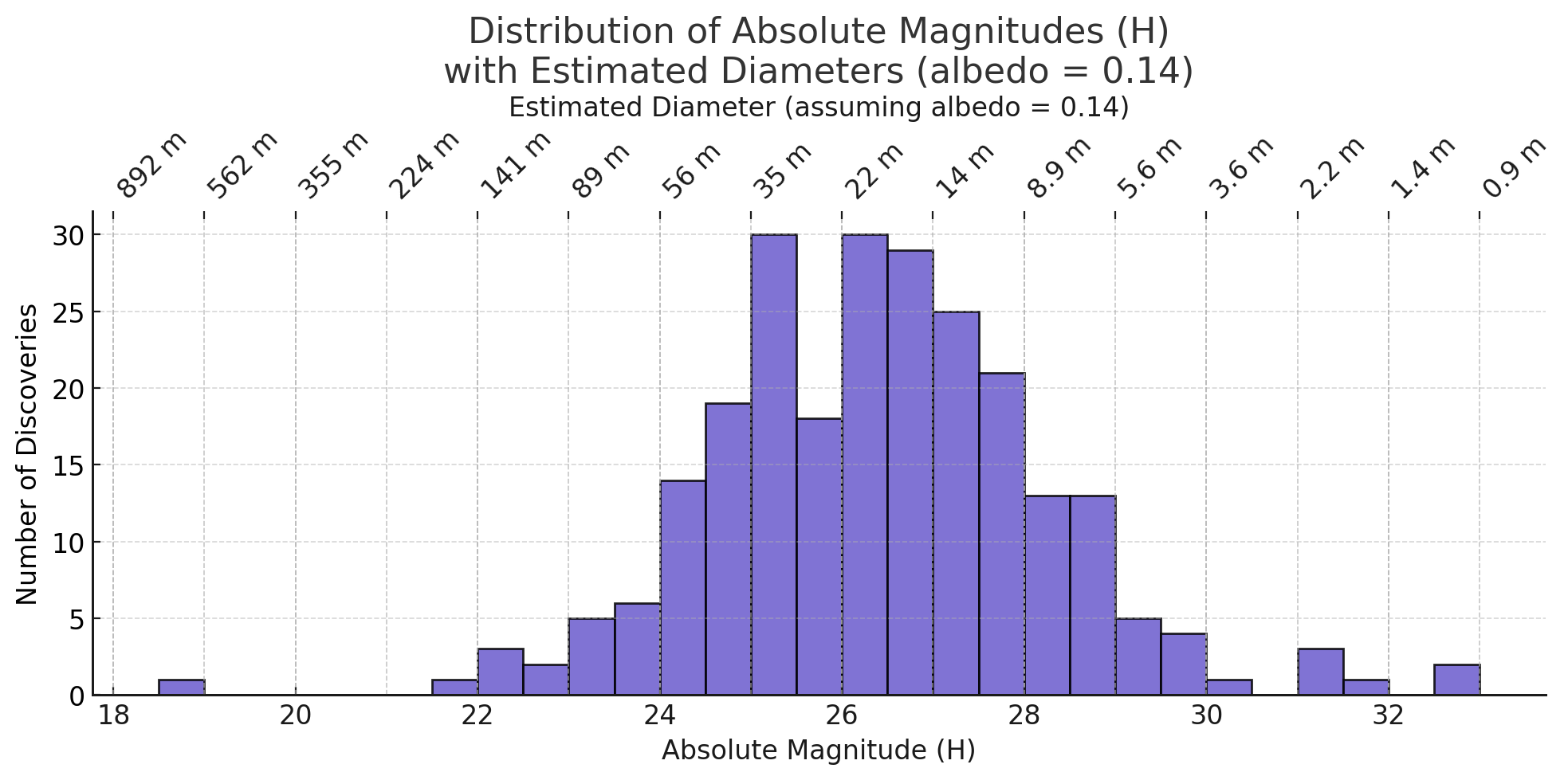}
    \caption{The absolute magnitude H distribution of the Piszk\'estet\H{o} NEOs between 2022 and 2024. The three imminent impactors all belong to the faint end of the histogram (H$>$30 mag).}
    \label{fig:absolute-mag}
\end{figure}

A physically more meaningful distribution is that of the absolute magnitude H, which can be converted into estimated diameters under different albedo assumptions. Assuming a typical NEO albedo of 0.14\cite{mainzer_NEO_albedo}, our survey is most sensitive to objects with $24 \leq H \leq 29$, corresponding to sizes between approximately 56 and 5.6 meters. The faintest objects in our dataset fall into the meter-sized regime, which represents the transitional domain between small asteroids and large meteoroids. This size range is of particular interest in current planetary science, as it bridges the populations of meteorite-producing meteoroids and NEOs tracked by dedicated surveys\cite{broz2024nature,broz2024aa,burdanov2025nature}.

\begin{figure}[h!]
    \centering
    \includegraphics[width=.7\linewidth]{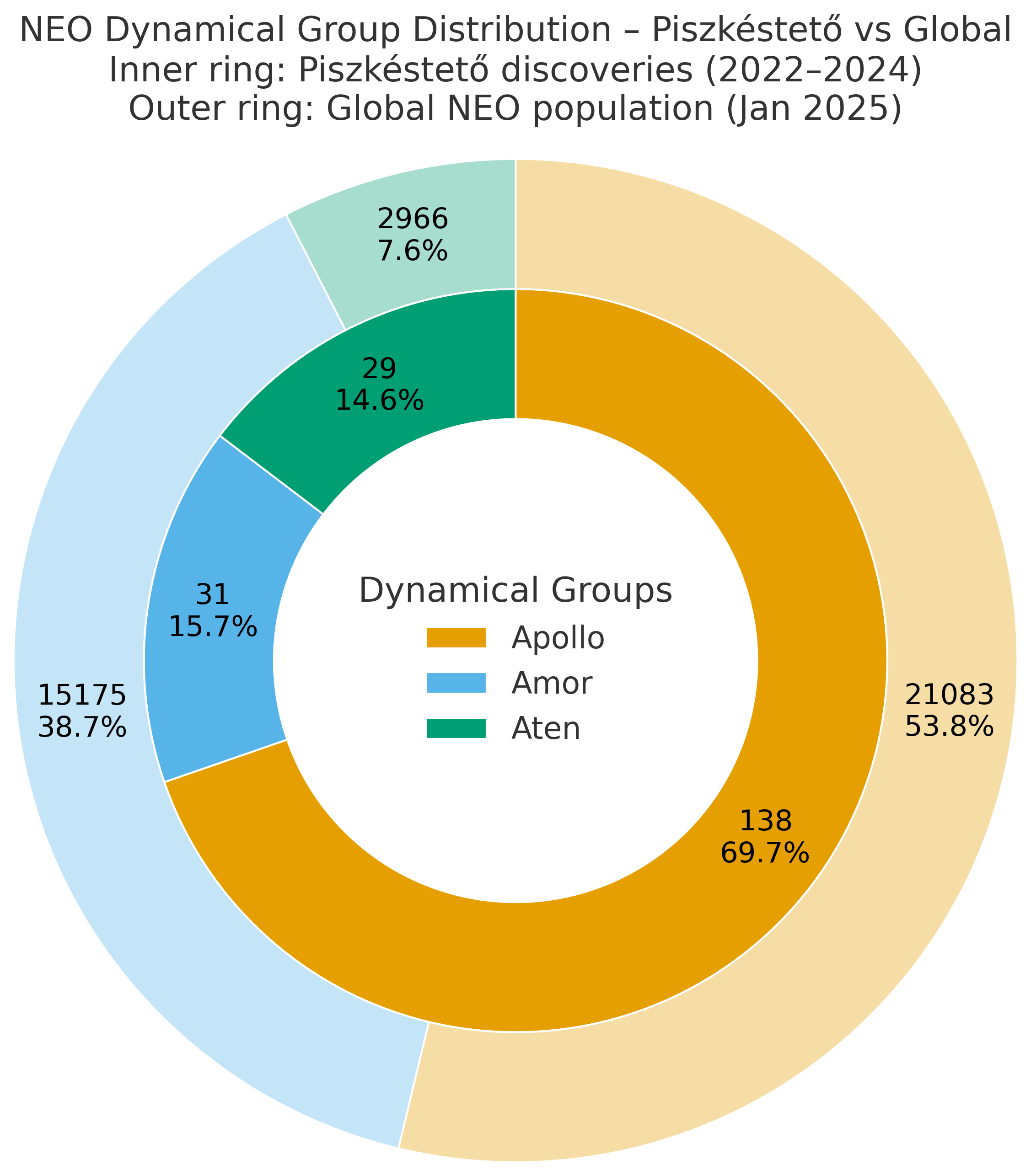}
    \caption{Comparison of NEO dynamical group distributions. The inner ring shows the relative proportions of NEOs discovered by the Piszkéstető survey between 2022 and 2024, while the outer ring represents the global NEO population as of January 2025.}
    \label{fig:asteroid_group}
\end{figure}

In addition to size and brightness, the orbital classification of the discovered objects offers further insight into the survey’s selectivity. Figure~\ref{fig:asteroid_group} compares the distribution of NEO discoveries by dynamical group at Piszkéstető with the global NEO population. The relative proportions for each group are as follows:
\begin{itemize}
\item \textbf{Apollo asteroids:} 138 discoveries (70\%) at Piszkéstető versus 21,083 globally (55\%).
\item \textbf{Amor asteroids:} 31 discoveries (16\%) at Piszkéstető versus 15,175 globally (40\%).
\item \textbf{Aten asteroids:} 29 discoveries (14\%) at Piszkéstető versus 2,966 globally (8\%).\footnote{\tt https://cneos.jpl.nasa.gov/stats/totals.html}
\end{itemize}

If our discoveries followed the same proportions as the global NEO population, we would expect approximately 109 Apollos, 79 Amors, and 16 Atens among the 198 Piszkéstető detections. Instead, the observed distribution shows a clear excess of Apollos and Atens and a deficit of Amors. The proportional comparison yields enhancement factors of about 1.27$\times$ for Apollos, 0.40$\times$ for Amors, and 1.75$\times$ for Atens relative to their global abundances. A chi-square goodness-of-fit test confirms that this deviation is highly significant ($\chi^{2}=40.7$, $p \ll 0.01$), indicating that the difference cannot be attributed to random sampling. This demonstrates that our survey is intrinsically more sensitive to Earth-crossing orbits, particularly those of Apollo-type asteroids, while Amors—remaining exterior to Earth’s orbit—spend less time in the sky regions covered most efficiently by our observing strategy. Atira-type asteroids, whose orbits lie entirely within that of Earth, are essentially undetectable by our survey, as observations are performed near opposition and never close to the Sun where such objects would appear. Such sensitivity toward Earth-crossing orbits is especially relevant for planetary-defense objectives, as it increases the likelihood of detecting objects on potentially impacting trajectories.

Apollo-type asteroids are Earth-crossing and include many of the objects most likely to cause significant atmospheric or surface events. The Chelyabinsk meteor of 2013, which caused widespread damage and injuries in Russia, is believed to have originated from an Apollo-type asteroid.\cite{Chelyabinsk2023} This connection reinforces the practical value of our current sensitivity range, as our survey is best suited to detect objects of similar type and size before impact.

In conclusion, our survey is mostly sensitive to possible Chelyabinsk-size impactors or ever smaller objects, thus providing a valuable service to the global efforts in planetary defense activities.

%% file: s-sec3.tex
\section{Highlights from Piszk\'estet\H{o}: three imminent impactors so far} 

\noindent The Piszk\'estet\H{o} NEO survey led to a surprising series of imminent impactor discoveries in the past three years. Between 2022 and 2024 we identified three objects -- 2022 EB5, 2023 CX1, and 2024 BX1 -- which were successfully tracked prior to atmospheric entry. Below, we summarize the discovery circumstances and follow-up observations for each case. 

Table \ref{tab:impact_summary} summarizes the most relevant observational and physical parameters for the three imminent impactors discovered from Piszkéstető.

\medskip

\noindent {\bf 2022 EB5:} During a routine survey on 11 March 2022, KS discovered the moderately bright object 2022 EB5 at 19:24 UTC, when it was approximately 109,000 km from Earth and about two hours prior to atmospheric entry. The asteroid had an apparent magnitude of 17.6–17.7 at discovery, and the finding was reported to the Minor Planet Center (MPC) approximately eight minutes later.

\begin{figure}
    \centering
    \includegraphics[width=0.9\linewidth]{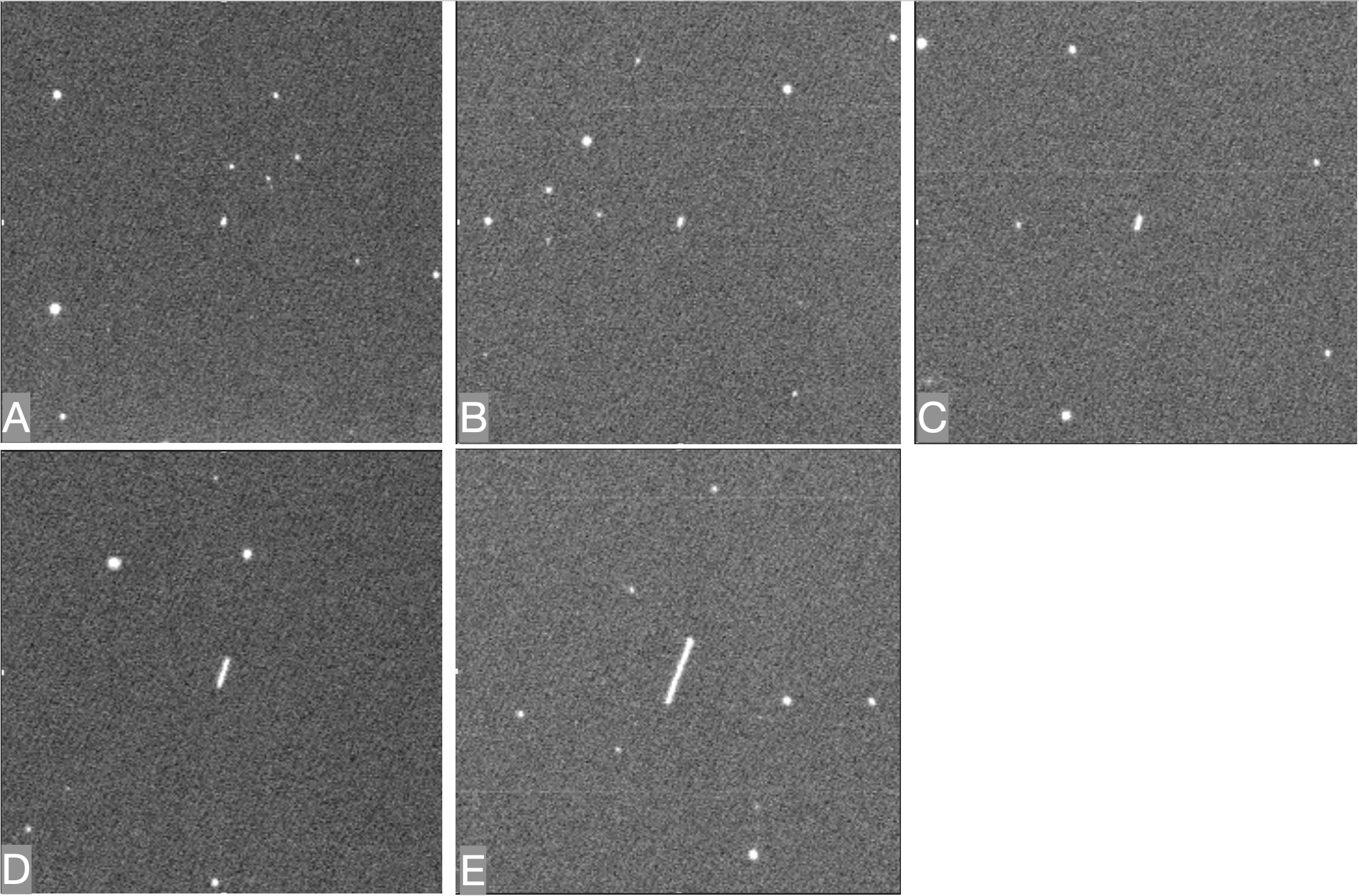}
    \caption{To highlight how the approaching asteroid was accelerating on the sky, here we show five 1 sec exposures of 2022 EB5 on 11 March 2022. A -- 20:44 UT; B -- 20:50 UT; C -- 20:55 UT; D -- 21:05 UT; E -- 21:11 UT. The last one was taken only 11 minutes before the atmospheric entry. }
    \label{fig:fig7}
\end{figure}

Follow-up observations were obtained between 19:51 and 19:54 UTC, by which time the asteroid's apparent motion had accelerated from an initial 52$^{\prime\prime}$/minute to 86–87$^{\prime\prime}$/minute,  an Earth-bound orbit (which was initially believed) or a likely impact trajectory. Astrometric measurements were submitted to the MPC after this second imaging series at 20:09 UTC. Continued tracking efforts allowed the asteroid to be observed until 11 minutes before impact, when it was only about 11,000 km from Earth's center (Fig. \ref{fig:fig7}). Scout from NASA Jet Propulsion Laboratory and Meerkat  from the European Space Agency both identified the object as an imminent impactor.\footnote{Observations were also made by the following observatories: 104, 246, 595, G02, N86, N88 (142/178 obs).}

The Center for Near-Earth Object Studies (CNEOS) at NASA later detected the visible radiation from the object's atmospheric entry over the Norwegian Sea, near coordinates 70.0° N, 9.1° W. Post-impact analysis suggests that 2022 EB5 was a C-type asteroid, with an estimated diameter of 5–6 meters, a comet-like density, and a very low albedo of less than 0.025 \cite{2022EB5_prop}.

\noindent {\bf 2023 CX1:} Less than a year later, KS discovered the asteroid 2023 CX1 on 12 February 2023 at 20:18:07 UTC. At the time of discovery, the object was already within the orbit of the Moon, at a distance of only 0.61 lunar distances (approximately 233,000 km), and had an apparent magnitude of 19.4. It was moving rapidly across the northern sky with an angular rate of 14$^{\prime\prime}$/minute and an inbound radial velocity 
of 9 km/s.

Recognizing its near-Earth nature immediately, KS assigned the temporary designation Sar2667 and reported it to the MPC's Near-Earth Object Confirmation Page (NEOCP) at 20:49 UTC, requesting urgent follow-up observations. A second observation half an hour later revealed that the asteroid was on a collision course with Earth. Subsequent observations by the Vi\v{s}njan Observatory in Croatia, beginning at 21:03 UTC, confirmed the imminent impact.

JPL's Scout system flagged the object as an imminent impactor and the European Space Agency promptly publicized the event via social media, and observatories worldwide contributed to additional observations\footnote{Observations were also made by the following observatories: 104, 203, 204, 587, 703, 958, A17, C65, C82, C95, D03, G18, I93, K63, K91, L01, L54, M26, M29, V39, Y66 (362/368 obs).}, significantly refining the predicted impact trajectory. 2023 CX1 reached a peak brightness of magnitude 13 shortly before entering Earth's shadow around 02:50 UTC on 13 February. It then faded rapidly and became unobservable until the impact. The final observation was recorded by Jost Jahn at the SATINO Remote Observatory in Haute Provence, France, at 02:52:07 UTC—only seven minutes before impact—when the asteroid was approximately 11,100 km from Earth's center (equivalent to about 4,700 km above Earth's surface).

The MPC assigned the official provisional designation 2023 CX1 at 04:13 UTC, approximately one hour post-impact. In total, at least 20 observatories submitted over 300 astrometric measurements before impact.

2023 CX1 was the seventh asteroid discovered prior to its impact with Earth, and notably, the third from which meteorite fragments were successfully recovered\cite{cx1_nature}. It marked KS’s second discovery of an impacting asteroid, following 2022 EB5.

\noindent {\bf 2024 BX1:} Continuing the sequence of pre-impact discoveries, the very small asteroid 2024 BX1, estimated to be only 40–50 cm in diameter\cite{2024BX1_aubrite}, was detected on 20 January 2024 at 21:48 UTC \cite{2024BX1_disc}, less than three hours before atmospheric entry. The discovery was made in a series of twelve 11-second exposures, during which the asteroid exhibited an apparent magnitude of 18.0 and a proper motion of 16$^{\prime\prime}$/minute. At the time of detection, it was located approximately 112,000 km from Earth.

Coordinates were reported to the MPC about 10 minutes after the initial observations, and follow-up imaging began 19 minutes later. Within the next 27 minutes, three follow-up observations were posted to the NEOCP. Both NASA and ESA warning systems quickly identified the object as a possible impactor.\textbf{\textit{\footnote{Observations were also made by the following observatories: 033, 073, 104, 203, 958, G18, G34, I93, K51, K63, L01, M38, Z21, Z31 (321/371 obs).}}}

A second series of observations commenced at 22:30 UTC, by which time the asteroid had brightened by 0.5 magnitudes compared to the time of discovery. Approximately 70 minutes after the initial detection, warning systems reported a 100\% probability of impact, predicting atmospheric entry over Germany, approximately 60 km west of Berlin, at 00:33 UTC on 21 January 2024.

The final observation of 2024 BX1 was obtained at 00:24:44 UTC, just eight minutes before impact, when the object was only about 6,500 km from Earth's surface.

The recovered meteorites from 2024 BX1, officially named Ribbeck, were classified as aubrites \cite{bischoff2024aubrite}. As discussed in the Introduction, aubrites are linked to enstatite-rich (E-type) asteroids, and it is notable that one of their proposed parent bodies, the Near-Earth Asteroid (3103) Eger \cite{Eger_aubrite}, was discovered by Mikl\'os Lovas using the same 0.6-meter Schmidt-telescope at Piszk\'estet\H{o} that KS later used to discover 2024 BX1.

\textbf{\textit{\begin{table}[ht]
\caption{Summary of key parameters for the three imminent impactors discovered from Piszkéstető. Distance, motion and magnitude values are taken at the time of discovery.}
\label{tab:impact_summary}
\centering
\begin{tabular}{lccccccc}
\hline
\textbf{Object} & \textbf{Dist. [km]} & \textbf{[LD]} & \textbf{Motion [$''$/min]} & \textbf{Time to Impact} & \textbf{Est. Size} & \textbf{Mag} & \textbf{Obs (Stations)} \\
\hline
2022 EB\textsubscript{5}  & 109,000 & 0.28 & 52 & $\sim$2 h     & 5–6 m     & 17.6–17.7 & 142 (6)   \\
2023 CX\textsubscript{1}  & 233,000 & 0.61 & 14 & $\sim$6.5 h   & $72\pm6$ cm\cite{cx1_nature} & 19.4      & 362 (20)  \\
2024 BX\textsubscript{1}  & 112,000 & 0.29 & 16 & $\sim$2.75 h  & 0.4–0.5 m & 18.0      & 371 (14)  \\
\hline
\end{tabular}
\end{table}}}

In all these discoveries --- apart from luck --- we think the most critical factor was real-time data processing. As soon as a NEO candidate is found (within minutes of the initial exposures), we stop the survey and switch to follow-up observations until we secure enough precision for impact probability determination and for orbit predictions for other follow-up observers. At this stage this is done by human interaction and is driven by the fact that there is a single dedicated telescope for NEOs at the Piszk\'estet\H{o} station.

%% file: s-conc.tex
\section{Conclusions and further outlook} 

\noindent With the recent CCD upgrade, we have reached the limits of traditional techniques. That is why we are now entering a new phase that aims at real-time discovery of NEOs via accelerated image analysis with AI methods. These efforts and a new web-based service we developed for other NEO surveys are described in a companion paper by \cite{piszkes_neo_ai}. 

Given the current approach, namely that when a NEO candidate is identified during the night, the Schmidt-telescope immediately switches from search to follow-up mode, it is clear that there is an increasingly important need for at least a second, fully dedicated telescope on-site that is capable both of searching and doing the follow-up for fresh targets. We are currently in the process of seeking funding for this second instrument, which we envision to essentially double the discovery rate from our Piszk\'estet\H{o} station. 

While placing a second telescope at a different site within Hungary is also under consideration, which would provide independent weather conditions and valuable parallactic information for close-approaching NEOs, this strategy comes with a drawback: if a discovery is made but conditions at the remote site do not permit follow-up, we would still need to interrupt the Schmidt telescope’s survey activity. For this reason, an on-site solution remains the most efficient option for maximizing our discovery yield.

With NEOs we have the advantage that no comparable survey exists at our geographic longitude, a strength that we would like to utilize as long as we can. Large upcoming facilities such as the Vera C. Rubin Observatory (LSST) will yield important results for NEO surveys in general, but their contribution to imminent impactor discoveries will remain limited. Our targets often become detectable only a few hours prior to impact, at apparent magnitudes and sky locations that are unfavorable for wide-field surveys like LSST, especially at the northern declinations accessible from Piszkéstető. For such very short-warning events, the geographic longitude of the observatory and the availability of real-time data processing remain decisive factors: if a telescope is not operating at the right place at the right time, the opportunity is lost. This is why smaller, longitude-distributed surveys continue to play a vital role in planetary defense alongside large international efforts.
\vspace{1cm}

\noindent \textbf{CRediT authorship contribution statement}

\textbf{Norton O. Szabó:} Writing – original draft, Visualization, Formal
analysis. \textbf{Krisztián Sárneczky:} Validation, Supervision, Resources, In-
vestigation. \textbf{László L. Kiss: Writing – original draft, Funding acquisi-
tion.} \textbf{Szabolcs Velkei:} Software, Data curation.

%% file: ack.tex
\section*{Acknowledgments}

\noindent The NEO survey at the Piszk\'estet\H{o} Mountain Station has been supported by a great variety of projects over the years. The latest instrument upgrade was funded by the GINOP-2.3.2-15-2016-00003 grant of the Hungarian National Research, Development and Innovation Office (NKFIH). The current activities are supported by the NEODetect project of the European Space Agency (ESA Contract No. 4000144612/24/D/BL). BS is supported by the NKFIH Élvonal grant KKP-143986. NOSZ is supported by the EK\"OP-25 university research scholarship program of the Ministry for Culture and Innovation from the source of the National Research, Development and
Innovation Fund and by the undergraduate research assistant program
of Konkoly Observatory. We sincerely thank the anonymous reviewers for their detailed and constructive feedback. Their suggestions -- from the inclusion of additional references and background information, to clarifications of our methods and the presentation of key results -- have greatly improved the quality and readability of this manuscript.
\vspace{1cm}

\noindent \textbf{Declaration of competing interest}

The authors declare the following financial interests/personal rela-
tionships which may be considered as potential competing interests:
Laszlo L. Kiss reports financial support was provided by European
Space Agency. If there are other authors, they declare that they have
no known competing financial interests or personal relationships that
could have appeared to influence the work reported in this paper.